\documentclass[twocolumn,secnumarabic,amssymb, nobibnotes, aps]{revtex4-1}
\usepackage{graphicx}
\usepackage{dcolumn}
\usepackage{bm}
\begin{document}

\title{Spin-resolved Andreev levels and parity crossings in hybrid superconductor-semiconductor nanostructures}%

\author{Eduardo J. H. Lee$^{1}$}
\author{Xiaocheng Jiang$^{2}$}
\author{Manuel Houzet$^{1}$}
\author{Ram\'{o}n Aguado$^{3}$}
\author{Charles M. Lieber$^{2}$}
\author{Silvano De Franceschi$^{1}$}
\email{silvano.defranceschi@cea.fr}

\affiliation{$^{1}$SPSMS, CEA-INAC/UJF-Grenoble 1, 17 rue des Martyrs, 38054 Grenoble Cedex 9, France}
\affiliation{$^{2}$Harvard University, Department of Chemistry and Chemical Biology, Cambridge, MA, 02138, USA}
\affiliation{$^{3}$Instituto de Ciencia de Materiales de Madrid (ICMM), Consejo Superior de Investigaciones Cient\'{i}ficas (CSIC), Sor Juana In\'{e}s de la Cruz 3, 28049 Madrid, Spain}

\begin{abstract}
{Updated version of this manuscript is available at Nature Nanotech. \textbf{9} (2014) 79.}
\end{abstract}

\date{\today}%

\maketitle


\textbf{
The hybrid combination of superconductors and low-dimensional semiconductors offers a versatile ground for novel device concepts \cite{Sdfreview}, such as sources of spin-entangled electrons \cite{Hofstetter, Herrmann, DasCooper}, nano-scale superconducting magnetometers \cite{Cleuziou}, or recently proposed qubits based on topologically protected Majorana fermions \cite{Sau,Lutchyn,vonOppen}. The underlying physics 
behind such hybrid devices ultimately rely on the magnetic properties of sub-gap excitations, known as Andreev levels.  
Here we report the Zeeman effect on the Andreev levels of a semiconductor nanowire quantum dot (QD) 
strongly coupled to a conventional superconductor. 
The combination of the large QD $g$-factor with the large superconductor critical magnetic field allows spin degeneracy to be lifted without suppressing superconductivity. We show that a Zeeman-split Andreev level crossing the Fermi energy signals a quantum phase transition in the ground state of the superconductivity-induced QD, denoting a change in the fermionic parity of the system. This transition manifests itself as a zero-bias conductance anomaly appearing at a finite magnetic field, with properties that resemble those expected for Majorana fermions in a topological superconductor \cite{MajoDelft,MajoWeizmann,MajoLund,Finck,Churchill}. Although the herein reported zero-bias anomalies do not hold any relation with topological superconductivity, the observed parity transitions can be regarded as precursors of Majorana modes in the long-wire limit \cite{Stanescu}. 
}


\begin{figure} [h]
\includegraphics[width=86mm]{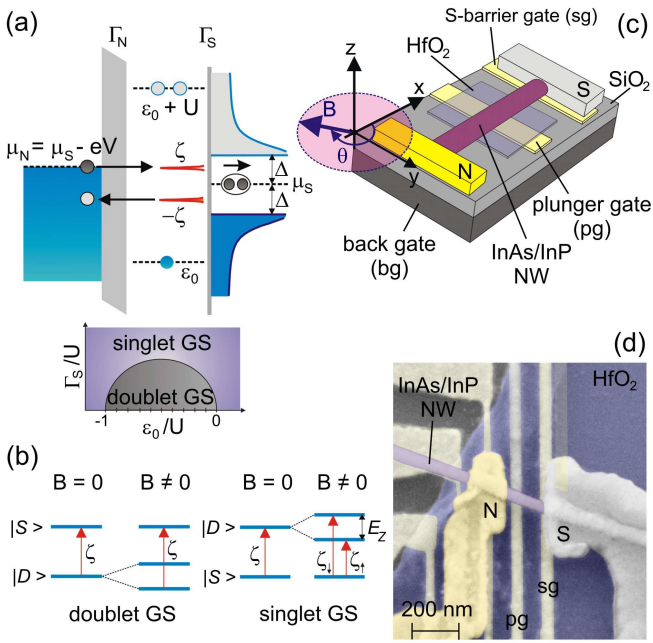}
\caption{\label{fig:epsart}\textbf{Andreev levels in a hybrid $S$-QD-$N$ system and device description.} (a) (Upper panel) Schematics of a $S$-QD-$N$ device with asymmetric tunnel couplings to the normal metal (Au) and superconductor (V) leads, $\Gamma_{N}$ and $\Gamma_{S}$, respectively. $\Delta$ is the superconducting gap, $U$ is the charging energy, $\mu_{i}$ is the chemical potential of the $i$ lead, and $\epsilon_{0}$ is the QD energy level relative to $\mu_S$ (in the $\Gamma_S \rightarrow 0$ limit, the QD has 0, 1 or 2 electrons for $\epsilon_0 > 0$, $-U<\epsilon_0<0$, $\epsilon_0<-U$, respectively). The sub-gap peaks located at $\pm\zeta$ represent the Andreev levels. In tunnel spectroscopy measurements the alignment of $\mu_N$ to an Andreev level yields a peak in the differential conductance. This process involves an Andreev reflection at the QD-$S$ interface, transporting a Cooper pair to the $S$ lead and reflecting a hole to the $N$ contact. (Lower panel) Qualitative phase diagram \cite{Rozhkov99, Vecino03, Simon} depicting the stability of the magnetic doublet ($|D\rangle=|\uparrow\rangle$, $|\downarrow\rangle$) versus that of the BCS singlet ($|-\rangle$). 
(b) Low-energy excitations of the QD-$S$ system and their expected evolution in a magnetic field, $B$. Doublet GS case (left): $|\uparrow\rangle$ is stabilized by $B$ and Andreev levels related to the transition $|\uparrow\rangle \rightarrow |-\rangle$ are observed. Singlet GS case (right): at finite $B$, the excited spin-split states $|\uparrow\rangle$ and $|\downarrow\rangle$ give rise to distinct Andreev levels with energy $\zeta_{\uparrow}$ and $\zeta_{\downarrow}$, respectively. $E_Z = |g|\mu_B B$ is the Zeeman energy, where $|g|$ is the $g$-factor and $\mu_B$ is the Bohr magneton. (c) Device schematic. The $N$ and $S$ leads were made of Ti(2.5 nm)/Au (50 nm) and Ti(2.5 nm)/V (45 nm)/Al (5 nm), respectively. The QD system is tuned by means of three gates: a plunger gate (pg), a barrier gate (sg) close to the $S$ contact, and a back gate (bg). $B$ is applied in the ($x,y$) device plane ($x$ being parallel to the NW). (d) Scanning electron micrograph of a $S$-QD-$N$ device.
}
\end{figure}

\begin{figure*}
\includegraphics[width=160mm]{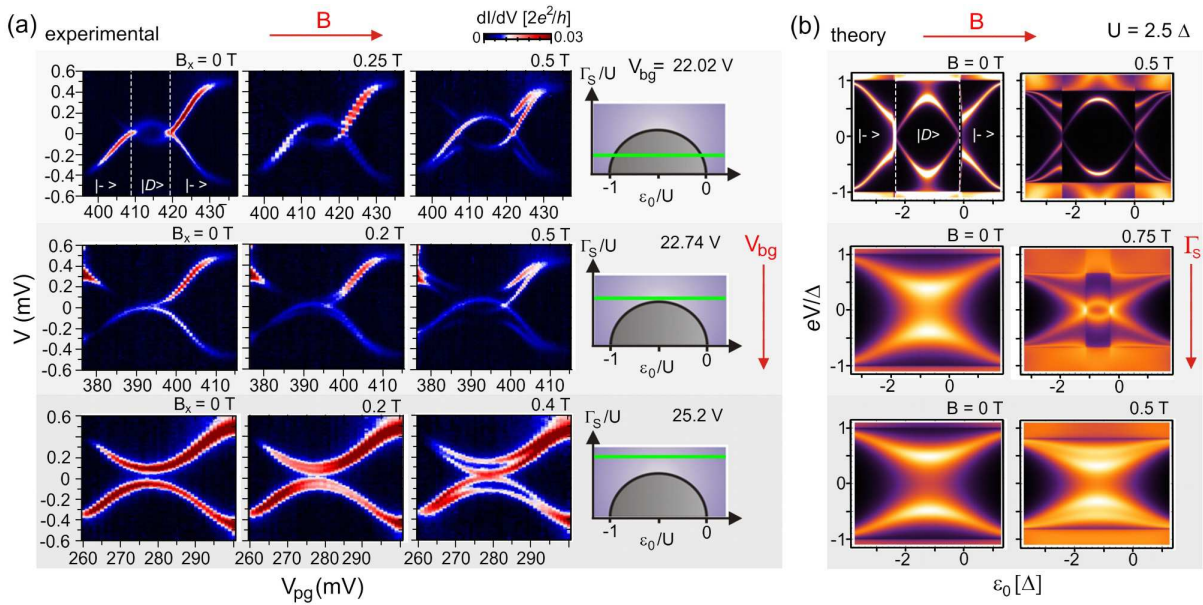}
\caption{\label{fig:epsart}\textbf{Andreev levels in different coupling regimes and their magnetic-field dependence.} (a) Experimental plots of $dI/dV$ vs. $(V_{pg},V)$ for different QD-$S$ couplings, $\Gamma_S$ (increasing from top to bottom), and different $B$ values (increasing from left to right). Top-left panel: along the $V_{pg}$ range, the system GS changes from singlet ($|-\rangle$) to doublet ($|D\rangle$) and back to singlet, following the green trajectory in the qualitative diagram on the right side of the same row. We find that increasing $V_{bg}$ results in larger $\Gamma_S$, thereby leading to an upward shift in the phase diagram. Eventually, the green trajectory is pushed into the singlet region (mid and bottom diagrams). Experimentally, this results in the disappearance of the doublet GS loop structure, as shown in the mid-left and bottom-left panels. The middle and right columns show the $B$-evolution of the Andreev levels in the three coupling regimes. For relatively weak coupling (top row), the Andreev levels for a singlet GS split due to the Zeeman effect, whereas those for a doublet GS simply move apart. At intermediate coupling (middle row), $B$ induces a quantum-phase transition from a singlet to a spin-polarized GS, as denoted by the appearance of a loop structure (right panel). At the largest coupling (bottom row), the Zeeman splitting of the Andreev levels is clearly visible all over the $V_{pg}$ range. The splitting is gate-dependent with a maximum in the central region. (b) Theoretical $dI/dV$ plots calculated by means of a self-consistent Hartree-Fock theory. In all calculations we used $U = 2.5\Delta$ and $\Gamma_N/\Gamma_S$ = 1/3. From top to bottom, $\Gamma_S$ was set to 0.2$\Delta$, 0.7$\Delta$ and 0.9$\Delta$. In spite of the relative simplicity of the Anderson model, the full experimental phenomenology is recovered.
}
\end{figure*}

\begin{figure*}
\includegraphics[width=160mm]{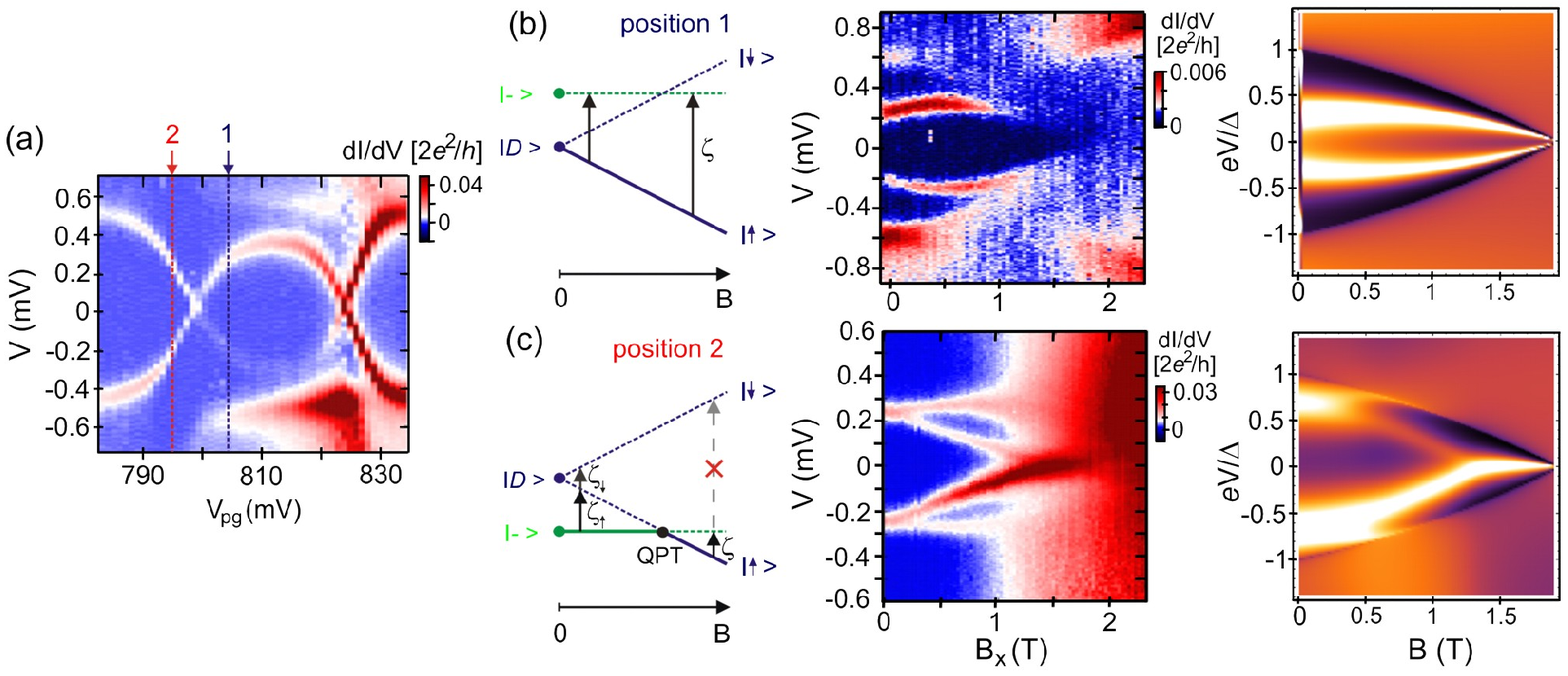}
\caption{\label{fig:epsart}\textbf{Magnetic-field evolution of the Andreev levels at fixed gate voltage and the level repulsion effect.} (a) $dI/dV(V_{pg},V)$ measurement at $B=0$ corresponding to a singlet-doublet-singlet sweep.(b) Left panel: Qualitative $B$-evolution of the low-energy states of a QD-$S$ system as expected for a doublet GS. Middle panel: Corresponding experimental data measured at position 1 in (a). $\zeta$ increases linearly with $B$ until it approaches the edge of the superconducting gap. The levels then move towards zero following the $B$ suppression of $\Delta$. Right panel: Corresponding theory plot of $dI/dV(B,eV/\Delta)$ calculated using experimentally measured parameters, i.e., $U/\Delta = 8$, $\epsilon_0/\Delta = -4$, and $\Gamma_S/\Delta = 1.5$ ($\Gamma_N/\Gamma_S$ was set to 1/3). (c) same as (b), but for singlet GS. The experimental plot in the middle panel was taken at position 2 in (a). It shows an asymmetric splitting of the $\zeta_{\uparrow}$ and $\zeta_{\downarrow}$ peaks. The weak $B$ dependence of $\zeta_{\downarrow}$ is due to the level repulsion between $|\downarrow\rangle$ and the continuum of quasiparticle states above $\Delta$. The corresponding numerical results were taken with the same parameters as in (b), except $\epsilon_0/\Delta = -6$.
}
\end{figure*}

When a normal-type ($N$) conductor is connected to a superconductor ($S$), superconducting order can leak into it giving rise to pairing correlations and to an induced superconducting gap. This phenomenon, known as the superconducting proximity effect, is expected also when the $N$ conductor is reduced to a small quantum dot (QD) with a discrete electronic spectrum. In this case, the superconducting proximity effect competes with the Coulomb blockade phenomenon, which follows from the electrostatic repulsion among the electrons of the QD. While superconductivity priviliges the tunneling of electron pairs with opposite spin, thereby favoring QD states with even numbers of electrons and zero total spin (i.e. spin singlets), the local Coulomb repulsion enforces a one-by-one filling of the QD, thereby stabilizing not only even but also odd electron numbers.  

In order to analyse this competition, let us consider the elementary case of a QD with a single, spin-degenerate orbital level. When the QD is singly-occupied, two ground states (GSs) are possible: 
a spin-doublet ($S = 1/2$), $|D\rangle = |\uparrow\rangle, |\downarrow\rangle$, or a spin singlet ($S = 0$), $|S\rangle$, whose nature has two limiting cases. 
In the large superconducting gap limit ($\Delta \rightarrow \infty$), the singlet is superconducting-like, $|S\rangle = -v^{*}|\uparrow\downarrow\rangle + u|0\rangle$, corresponding to a Bogoliubov-type superposition of the empty, $|0\rangle$, and doubly-occupied, $|\uparrow\downarrow\rangle$, states. 
In the strong coupling limit, where the QD-S tunnel coupling, $\Gamma_S$, is much larger than $\Delta$, quasiparticles in the superconductor screen the local magnetic moment in the QD and the singlet is Kondo-like. Although the precise boundary between these limiting cases is not well-defined \cite{Yamada}, it is possible to unambiguously detect changes in the parity of the ground state of the system, i.e.,  whether the GS is a singlet (even fermionic parity) or a doublet (odd fermionic parity), as we show here. The competition between the singlet and doublet states is determined by different energy scales: $\Delta$, $\Gamma_S$, the charging  energy, $U$, and the energy $\epsilon_0$ of the QD level relative to the Fermi energy of the $S$ electrode (see Fig. 1a) \cite{Glazman, Rozhkov99, Vecino03, Hewson04,Hewson07, Choi, Simon, Domanski, Koenig2013}. 
Previous works addressing this competition have focused either on the Josephson current in S-QD-S devices \cite{sdf2006, Lindelof, Wernsdorfer12} or on the sub-gap structure in S-QD-S and N-QD-S devices \cite{Buitelaar, Kanai, Jesp, Eichler, Pillet, Deacon2010, Mason, Lee2012, Chang, Pillet2013, Kumar}.  


Here we investigate the magnetic properties of the lowest-energy 
states in a $S$-QD-$N$ geometry, where the $N$ contact acts as a weakly coupled tunnel probe. In this geometry, a direct spectroscopy of the density of states (DOS) in the QD-$S$ system can be performed through a measurement of the differential conductance, $dI/dV$, as a function of the voltage difference, $V$, between $N$ and $S$. In such a measurement, an electrical current measured for $|V| < \Delta/e$ is carried by so-called Andreev reflection processes, each of which involves two single-electron transitions in the QD.  For example, an electron entering the QD from $N$ induces a single-electron transition from the QD GS, i.e. $|D\rangle$ or $|-\rangle$,  to the first excited state (ES), i.e. $|-\rangle$ or $|D\rangle$, respectively. The ES relaxes back to the GS through the emission of an electron pair into the superconducting condensate of $S$ and a second single-electron transition corresponding to the injection of another electron from $N$ (the latter process is usually seen as the retroreflection of a hole into the Fermi sea of $N$). The just described transport cycle yields a $dI/dV$ resonance, i.e. an Andreev level, at $eV = \zeta$, where $\zeta$ is the energy difference between ES and the GS, i.e. between $|D\rangle$ or $|-\rangle$, or vice-versa (see Fig. 1a). The reverse cycle, which involves the same excitations, occurs at $eV = - \zeta$, yielding a second Andreev level symmetrically positioned below the Fermi level. 

We used devices based on single InAs/InP core/shell nanowires (NWs), where vanadium (gold) was used for the $S$ ($N$) contact \cite{Giazotto}. A device schematic and a representative image are shown in Figs. 1c and 1d, respectively. The fabricated vanadium electrodes showed $\Delta = 0.55$ meV and an in-plane critical magnetic field $B_c^x \approx 2$ T ($x$ $\parallel$ NW axis). The QD is naturally formed in the NW section between the $S$ and $N$ contacts. We find typical $U$ values of a few meV (i.e., $U/\Delta \approx 3-10$).
The QD properties are controlled by means of two bottom electrodes crossing the NW, labeled as plunger gate and $S$-barrier gate, and a back gate provided by the conducting Si substrate. 
To achieve the asymmetry condition $\Gamma_S \gg \Gamma_N$, the $S$-barrier gate was positively biased at $V_{sg} = 2$ V. We used the plunger gate voltage $V_{pg}$ to vary the charge on the QD, and the back-gate voltage $V_{bg}$ to finely tune the tunnel coupling. 

Figure 2a shows a series of $dI/dV (V_{pg},V)$ measurements for three different $\Gamma_S$. The top row refers to the weakest 
$\Gamma_S$.  In this case, the spanned $V_{pg}$ range corresponds to a horizontal path in the phase diagram that goes through the doublet GS region (see right diagram). Let us first consider the leftmost plot taken at magnetic field $B=0$. On the left and right sides of this plot, the QD lies deep inside the singlet GS regime. Here the doublet ES approaches the superconducting gap edge, yielding an Andreev-level energy $\zeta \approx \Delta$. By moving towards the central region, the two sub-gap resonances approach each other and cross at the singlet-doublet phase boundaries, where  $\zeta = 0$. 
In the doublet GS regime between the two crossings, the sub-gap resonances form a loop structure with $\zeta$ maximal at the electron-hole symmetry point.
Increasing $\Gamma_S$ corresponds to an upward shift in the phase diagram. 
The middle row in Fig. 2a refers to the case where $\Gamma_S$ is just large enough to stabilize 
the singlet GS over the full $V_{pg}$ range (see right diagram). 
At $B=0$, the Andreev levels approach the Fermi level without crossing it. 
A further increase in $\Gamma_S$ leads to a robust stabilization of the singlet GS (bottom row). At zero-field, the sub-gap resonances remain distant from each other coming to a minimal separation at the electron-hole symmetry point ($\epsilon_0=-U/2$). 
Similarly to the case of superconducting single-electron transistors \cite{Devoret}, the QD occupation increases with $V_{pg}$ in units of two 
without going through an intermediate odd state.   

We now turn to the effect of $B$ on the Andreev levels (middle and right columns in Fig. 2a).  Starting from the weak coupling case (top row), a field-induced splitting of the sub-gap resonances appears, yet only in correspondence of a singlet GS. This is due to the fact that these resonances involve excitations between states of different parity. For a singlet GS, the spin degeneracy of the doublet ES is lifted by the Zeeman effect resulting in two distinct excitations (see Fig. 1b).
By contrast, for a doublet GS, no sub-gap resonance stems from the $|\uparrow\rangle \rightarrow |\downarrow\rangle$ excitation, because these two states have the same (odd) number of electrons. The energy of the only visible Andreev level, associated with the $|\uparrow\rangle \rightarrow |-\rangle$ transition, increases with $B$. 
The formation of a loop structure in the rightmost panel of the middle row shows that a QPT from a singlet to a spin-polarized GS can be induced by $B$
when the starting $\zeta$ is sufficiently small.
In the bottom row, Zeeman-split Andreev levels can be seen all over the spanned $V_{pg}$ range. At $B_x=0.4$ T, the inner levels overlap at the Fermi level, indicating a degeneracy between the $|\uparrow\rangle$ and $|-\rangle$ states. 
In Fig. 2b we show theoretical $dI/dV$ plots of a $S$-QD-$N$ Anderson model calculated by means of self-consistent Hartree-Fock theory \cite{Rodero12} (see Suppl. Information). The full phenomenology explained above is recovered, supporting our interpretation in terms of spin-resolved Andreev levels and a QPT.

\begin{figure}
\includegraphics[width=86mm]{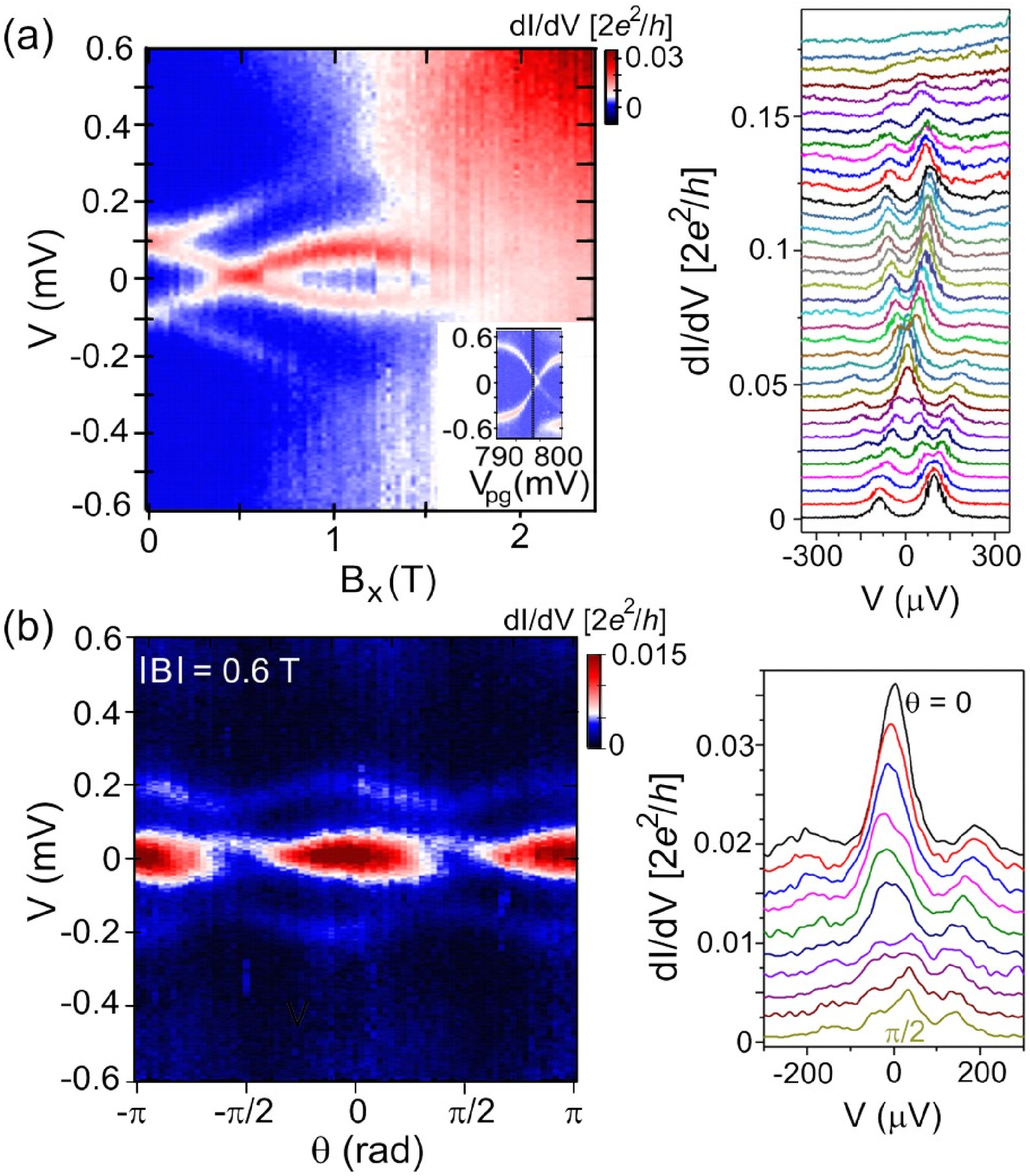}
\caption{\label{fig:epsart}\textbf{Magnetic-field induced QPT and angle anisotropy.} (a) Left panel: $dI/dV (B,V)$ taken at the position of the vertical line in the inset (same device as in Fig. 3). Right panel: line traces at equally spaced B values as extracted from the data in the left panel (each shifted by 0.005 $\times 2e^2/h$). The QPT induced by the field is observed as a ZBP extending over a $B$ range of about 150 mT. This apparently large extension is a consequence of the finite width of the Andreev levels. 
(b) $dI/dV(V)$ traces taken with $|B|$ = 0.6 T, at different angles. This field magnitude corresponds to the QPT field when $B$ is aligned to the NW axis at $\theta=0$ 
Owing to the $g$-factor anisotropy, the ZBP associated with the QPT is split and suppressed when $B$ is rotated away from the NW axis. The peak splitting has a maximum at $\theta=\pi/2$
}
\end{figure}

Interestingly, the splitting of Andreev levels appears to be gate dependent. It tends to vanish when the system is pushed deep into the singlet GS, and it is maximal near the phase boundaries. To further investigate this behaviour, we have measured $dI/dV(B,V)$ for fixed values of $V_{pg}$. 
These measurements were carried out on a second similar device (see Suppl. Information). 
The mid-panel of Fig. 3b displays the $B_x$ dependence of the sub-gap resonances measured at position 1 in Fig. 3a. Initially, the energy of the Andreev levels increases, as expected for a doublet GS (see left panel). From a linear fit of the low-field regime, i.e. $\zeta(B_x) = \zeta(0) +E_{Z}/2$, where $E_Z = |g_x|\mu_B B_x$ is the Zeeman energy and $\mu_B$ is the Bohr magneton, we obtain a $g$-factor $|g_x| \approx 5.6$. For $B_x >$ 0.7 T, the field-induced closing of the gap bends the Andreev levels down to zero-energy. Finally, above the critical field, a split Kondo resonance is observed, from which $|g_x| \approx$ 5.5 is estimated, consistent with the value extracted from the Andreev level splitting.  The mid-panel of Fig. 3c displays a similar measurement taken at position 2 in Fig. 3a, where the GS is a singlet. 
The splitting of the Andreev levels is clearly asymmetric. The lower level decreases to zero according to a linear dependence $\zeta_{\uparrow}(B_x) = \zeta(0)-E_{Z}/2$, with $|g_x| \approx$ 6.1, which is close to the value measured from the split Kondo resonance in the normal state. 
The higher energy level, however, exhibits a much weaker field dependence.
Both the non-linear field dependence for $B_x > 0.7$ T in Fig. 3b and the asymmetric splitting in Fig. 3c can be explained in terms of a level-repulsion effect between the Andreev levels and the continuum of quasiparticle states.
This interpretation is confirmed by numerical calculations shown in the right panels of Figs. 3b and 3c, which are in good agreement with the respective experimental data. In the mid-panel of Fig. 3c, the inner sub-gap resonances cross around 1.5 T, denoting a field-induced QPT. Above this field, however, they remain pinned as a zero-bias peak (ZBP) 
up to $B_c^x \approx 2$ T. This peculiar behavior, reproduced by the numerical results, can be attributed to the level-repulsion effect discussed above, in combination with the rapid shrinking of $\Delta$ with $B_x$. 

In order to observe a clear $B$-induced QPT from a singlet to a spin-polarized GS, we reduced $\zeta(0)$ by tuning $V_{pg}$ closer to the singlet-doublet crossing in Fig. 3a. The corresponding data are shown in Fig. 4a. Contrary to the case of Fig. 3c, the Andreev level splitting is rather symmetric, owing to the reduced importance of the level repulsion effect at energies far from $\Delta$. The inner sub-gap resonances split again after the QPT, which occurs now at $B_x \approx 0.5$ T. As expected, the outer sub-gap resonances get simultaneously suppressed (left panel of Fig. 3c). The suppression is not complete though, suggesting a partial population of the $|-\rangle$ ES, possibly favored by thermal activation.  

We note that the ZBP at the QPT appears to extend on a sizable field range $\Delta B_x \approx$ 150 mT. This range is consistent with the $\Gamma_{N}$-dominated lifetime broadening of the Andreev levels, i.e. $|g_x|\mu_{B}\Delta B_x \approx$ peak width $\approx 50 \mu$eV. In Fig. 4b we show how the ZBP depends on the in-plane $B$ angle, $\theta$, relative to the NW axis. 
As $\theta$ varies from 0 to $\pi/2$, the ZBP splits into two peaks with decreasing height. This angle dependence is an effect of the $g$-factor anisotropy. For $\theta = \pi/2$, we find a $g$-factor $|g_{y}| \approx 3$,  i.e. a factor of 2 smaller than for $\theta = 0$ (see Suppl. Information). As a result, the QPT only occurs at a higher field  (see Suppl. Information, $B_{QPT}^y \approx 1$ T), and the split peaks correspond to $\zeta_{\uparrow}$ transitions on the singlet-GS side. Figure 4b shows also a pair of small outer peaks associated with the $\zeta_{\downarrow}$ transitions. Their oscillatory position is as well due to $g$-factor anisotropy. 


Noteworthy, the $B$ dependences discussed above mimic some of the signatures expected for Majorana fermions in hybrid devices \cite{Lutchyn, vonOppen, Aguado12b, Loss12b, DasSarma12, Potter, MajoDelft, MajoWeizmann, MajoLund, Finck, Churchill}. A ZBP extending over a sizable $B$ range is observed for $\theta = 0$, and it is suppressed for $\theta = \pi/2$, i.e. when B is presumably aligned to the Rashba spin-orbit field, $B_{SO}$ \cite{MajoDelft, MajoWeizmann}. While in Fig. 4 the field extension of the ZBP is limited by the ratio between the Andreev-level linewidth and the $g$-factor, Fig. 3b shows a ZBP extending over a much larger $B$ range.  This stretching effect can be attributed to the field-induced suppression of $\Delta$ and the consequently enhanced level repulsion with the continuum of quasiparticle states. In larger QDs or extended nanowires, a similar level-repulsion effect may as well arise from other Andreev levels present inside the gap \cite{Aguado12b,Potter,Loss12b, Stanescu}. 


A more detailed discussion of the relation between the results presented here and existing experiments concerning Majorana fermions in hybrid devices is given in the Supplementary Information. Interestingly, a recent theoretical work has shown that zero-energy crossings of Andreev levels associated with a change in the ground state parity, similar to those presented here, adiabatically evolve towards zero-energy Majorana modes with increasing nanowire length \cite{Stanescu}. 
This evolution might be experimentally investigated by studying the $B$-evolution of Andreev levels in nanowires of increasing length. Along similar lines, recent proposals have discussed the possibility of exploring the short-to-long wire evolution in chains of magnetic impurities deposited on superconducting surfaces \cite{Nadj-Perge,KlinovajaRKKY, Franz2013, Braunecker, Pientka}. In such proposals, the Yu-Shiba-Rusinov bound states induced by the individual magnetic impurities (similar to the Andreev levels discussed here) may ultimately evolve towards Majorana modes localized at the edges of the atom chain, upon increasing the chain length.

\vspace{10 mm}

\textbf{Methods}

\textbf{Device fabrication}. The $S$-QD-$N$ devices used in this study were based on individual InAs/InP core/shell nanowires grown by thermal evaporation \cite{Xiaocheng} (diameter $\approx$ 30 nm, shell thickness $\approx$ 2 nm). The NWs were deposited onto Si/SiO$_2$ substrates on which arrays of thin metallic striplines [Ti(2.5 nm)/Au(15 nm), width $\approx$ 50 nm] covered by a 8 nm-thick atomic layer deposition (ALD) HfO$_2$ film had been previously processed. During the measurements, the degenerately-doped Si substrate was used as a global backgate, whereas the striplines were used as local gates. Normal metal [Ti(2.5 nm)/Au(50 nm)] and superconductor [Ti(2.5 nm)/V(45 nm)/Al(5 nm)] leads were incorporated to the devices by means of standard e-beam lithography techniques (lateral separation $\approx$ 200 nm). 

\vspace{10 mm}

This work was supported by the European Starting Grant program and by the Agence Nationale de la Recherche. R. A. acknowledges support from the Spanish
Ministry of Economy and Innovation through grants FIS2009-08744 and FIS2012-33521. The authors thank J.-D. Pillet for useful discussions.

\bibliography{lee2013}

\end{document}


\renewcommand{\figurename}{Fig. S}

\title{Supplementary Information: Probing the spin texture of sub-gap states in hybrid superconductor-semiconductor nanostructures}

\author{Eduardo J. H. Lee$^{1}$}
\author{Xiaocheng Jiang$^{2}$}
\author{Manuel Houzet$^{1}$}
\author{Ram\'{o}n Aguado$^{3}$}
\author{Charles M. Lieber$^{2}$}
\author{Silvano De Franceschi$^{1}$}

\affiliation{$^{1}$SPSMS, CEA-INAC/UJF-Grenoble 1, 17 rue des Martyrs, 38054 Grenoble Cedex 9, France}
\affiliation{$^{2}$Harvard University, Department of Chemistry and Chemical Biology, Cambridge, MA, 02138, USA}
\affiliation{$^{3}$Instituto de Ciencia de Materiales de Madrid (ICMM), Consejo Superior de Investigaciones Cient\'{i}ficas (CSIC), Sor Juana In\'{e}s de la Cruz 3, 28049 Madrid, Spain}

\maketitle

\section{Model and Hartree-Fock theory}
\label{model}

As described in the main text, our nanowires are well described by a single-level quantum dot with competing superconducting correlations and on-site Coulomb interactions.
A minimal model for describing this experimental configuration is given by a single level Anderson model coupled to a superconducting reservoir
with Hamiltonian $H_{QD-S}=H_S+H^S_T+H_{QD}$. Here, $H_{QD}$ models
the uncoupled quantum dot, and is given by
\begin{equation}
H_{QD} = \sum_{\sigma=\uparrow,\downarrow} \epsilon_\sigma d^{\dagger}_{\sigma} d_{\sigma} +
U n_{\uparrow} n_{\downarrow} ,
\label{HQD}
\end{equation}
where $d^{\dagger}_{\sigma}$ creates an electron with spin $\sigma$ on the dot level
located at $\epsilon_\sigma$. In the presence of an external magnetic field $B$, spin-degeneracy is broken and the levels are given by 
$\epsilon_\uparrow=\epsilon_0-\frac{1}{2}E_Z$ and $\epsilon_\downarrow=\epsilon_0+\frac{1}{2}E_Z$ with $E_Z=g\mu_B B$, being the Zeeman energy.
The second term in Eq. (\ref{HQD}) describes the local Coulomb interaction for two electrons
with opposite spin within the dot ($n_{\sigma}= d^{\dagger}_{\sigma}c_{\sigma}$), where $U$ is the charging energy.
$H_{S}$ describes the uncoupled superconducting lead, modelled 
by a BCS Hamiltonian
\begin{equation}
H_{S} = \sum_{k_S\sigma} \xi_{k_S} c^{\dagger}_{k_S\sigma} c_{k_S\sigma} +
\sum_{k_S} \left(\Delta c^{\dagger}_{k_S\uparrow}c^{\dagger}_{-k_S\downarrow} +
\mbox{h.c.} \right),
\label{Hnu}
\end{equation}
where $\xi_{k_S} = \epsilon_{k_S} - \mu_{S}$ is referred with respect to the chemical potential at the superconducting reservoir $\mu_{S}$
and $\Delta$ is the superconducting order parameter.
$H^S_T$ describes the coupling between the QD level and the superconductor and has the form
\begin{equation}
H^S_{T} = \sum_{k_S\sigma} \left(V_{k_S} c^{\dagger}_{k_S\sigma} d_{\sigma} + \mbox{h.c.} \right).
\nonumber
\label{HTS}
\end{equation}
This coupling to the superconducting lead is characterized by the parameter
$\Gamma_{S}=\pi\sum_{k_S} |V_{k_S}|^2\delta(\omega-\epsilon_{k_S})$. As described in the main text, the competition between the three energy scales $U$, $\Delta$ and $\Gamma_{S}$ governs the ground state of the model which can be either a singlet or a doublet.
Finally, the experimental setup includes a normal reservoir. We model this by adding two extra terms to the Hamiltonian such that the total model reads $H=H_{N} +H^N_T+H_{QD-S}$. The first term describes the normal reservoir and reads
$H_{N} = \sum_{k_N,\sigma} \xi_{k_N} c^{\dagger}_{k_N\sigma} c_{k_N\sigma}$,
where $c^{\dagger}_{k_N\sigma}$ creates an electron with spin $\sigma$ at the
single-particle energy level $\xi_{k_N} = \epsilon_{k_N} - \mu_{N}$, with $\mu_{N}$ being the chemical potential at the normal reservoir.
The coupling to the normal lead is given by the term 
\begin{equation}
H^N_{T} = \sum_{k_N\sigma} \left(V_{k_N} c^{\dagger}_{k_N\sigma} d_{\sigma} + \mbox{h.c.} \right),
\nonumber
\label{HTN}
\end{equation}
which defines $\Gamma_{N}=\pi\sum_{k_N} |V_{k_N}|^2\delta(\omega-\epsilon_{k_N})$.

All the relevant quantities for the experiment can be extracted from the
QD Green's functions in Nambu space defined as $\hat{G}^{r}_{\sigma}(t,t') = -i\theta(t-t') \langle \left[\Psi_{\sigma}(t)
,\Psi^{\dagger}_{\sigma}(t')\right]_+\rangle$, where 
$\Psi_{\sigma} = ( d_{\sigma} \; d^{\dagger}_{-\sigma})^T$.
In frequency space, the QD Green's function can be formally written as
$\hat{G}^{r}_{\sigma}(\omega)^{-1} = \hat{G}^{r,(0)}(\omega)^{-1}- 
\hat{\Sigma}_{\sigma}(\omega)$,
where $\hat{G}^{r,(0)}(\omega)$ is the non-interacting dot Green's function in Nambu
space and the self-energy $\hat{\Sigma}_{\sigma}$ takes into account both the coupling to the leads and the Coulomb interaction. Of course, the full problem cannot be exactly solved and one needs to resort to approximations. In the main text, we present calculations using a Hartree-Fock (HF) decoupling of the self-energy \cite{vecino}. It has been demonstrated that such HF decoupling gives reliable results when benchmarked against more sophisticated methods such as numerical renormalization group \cite{Martin-RoderoYeyaytiJPC}. The HF selfenergy is obtained by considering the first order diagrams in the Coulomb interaction. Its diagonal components in Nambu space are given by $\Sigma^{HF}_{11,\sigma} = -\Sigma^{HF}_{22,-\sigma}= U \langle n_{-\sigma}\rangle$, where the spin-resolved occupations $\langle n_{\sigma}\rangle=-\frac{1}{\pi}\int d\omega Im{G}^{r}_{11,\sigma}(\omega)$ have to be calculated \emph{self-consistently}. Importantly, also the anomalous self-energies
$\Sigma^{HF}_{12,\sigma} = (\Sigma^{HF}_{21,\sigma})^* = -U \langle d_{\uparrow} d_{\downarrow}\rangle$ have to be taken into account.
The explicit expression for $\hat{G}^{r,HF}_{\sigma} (\omega)$ reads
\begin{widetext}
\begin{equation}
\hat{G}^{r,HF}_{\sigma} (\omega)= \frac{1}{D(\omega)}\left( \begin{array}{cc} \omega +\epsilon_{-\sigma} +i\Gamma_N +\Gamma_S\frac{\omega}{\sqrt{\Delta^2-\omega^2}}
 + U \langle n_{\sigma}\rangle& 
 \Gamma_S\frac{\Delta}{\sqrt{\Delta^2-\omega^2}}+U \langle d_{\uparrow} d_{\downarrow}\rangle\\
 \Gamma_S\frac{\Delta}{\sqrt{\Delta^2-\omega^2}}+U \langle d_{\downarrow}^{\dagger} d_{\uparrow}^{\dagger}\rangle & 
\omega - \epsilon_\sigma +i\Gamma_N +\Gamma_S\frac{\omega}{\sqrt{\Delta^2-\omega^2}}
 - U \langle n_{-\sigma}\rangle 
\end{array} \right).
\label{hfa-GF}
\end{equation}
The Andreev level spectrum of the system is given by the roots of the determinant $D(\omega)\equiv Det[\hat{G}^{r,HF}_{\sigma} (\omega)^{-1}]$, namely by the solutions of
\begin{eqnarray}
(\omega - \epsilon_\sigma +i\Gamma_N +\Gamma_S\frac{\omega}{\sqrt{\Delta^2-\omega^2}}
 - U \langle n_{-\sigma}\rangle )(\omega +\epsilon_{-\sigma} +i\Gamma_N +\Gamma_S\frac{\omega}{\sqrt{\Delta^2-\omega^2}}
 + U \langle n_{\sigma}\rangle)\nonumber\\
-(\Gamma_S\frac{\Delta}{\sqrt{\Delta^2-\omega^2}}+U \langle d_{\uparrow} d_{\downarrow}\rangle)(\Gamma_S\frac{\Delta}{\sqrt{\Delta^2-\omega^2}}+U \langle d_{\downarrow}^{\dagger} d_{\uparrow}^{\dagger}\rangle)=0.
\label{andreev-level} 
\end{eqnarray}
\end{widetext}
The QD spectral function is defined as $A(\omega)=-\frac{1}{\pi}Im Tr[\hat{G}^{r,HF}_{\sigma}(\omega)]$, where the trace includes both spin and Nambu degrees of freedom.

In practice, the results presented in the main text were obtained by discretizing the Fourier space in a finite mesh of size $N=2^{18}$ with $\omega_i\in[-D,D]$ and cutoff $D=25\Delta$. Starting from an initial effective mean-field solution
$\Sigma_{11,\uparrow} = -\Sigma_{22,\downarrow}= \frac{U}{2}$ \cite{pillet2010}, we iterate the HF equations until good numerical convergence in the spin-resolved occupations is reached (as a criterium for good convergence, the iteration stops when the relative error between successive occupations in the iteration loop is less than $10^{-5}$).

For a given bias voltage $eV=\mu_{N}-\mu_{S}$, the conductance across the system is given by $\mathcal{G}=dI/dV$, where the total current can be decomposed into Andreev and quasiparticle contributions, $I=I_A+I_Q$. For voltages $eV\leq\Delta$, the quasiparticle current is zero, $I_Q=0$, and the only contribution comes from Andreev processes. The Andreev current reads
\begin{eqnarray}
I_A=\frac{2e}{h}\int d\omega[f_N(\omega-V)-f_N(\omega+V)]T_A(\omega),
\end{eqnarray}
where the Andreev transmission, defined as
\begin{equation}
T_A(\omega)=4\Gamma_N^2\sum_\sigma |G^{HF}_{12,\sigma}(\omega)|^2,
\end{equation}
describes processes in which an electron (hole) from the normal side is reflected as a hole (electron) while an extra Cooper pair is created on the superconducting side.
For voltages above the gap, also the quasiparticle contribution is finite and is given by:
\begin{eqnarray}
I_Q=\frac{2e}{h}\int d\omega[f_N(\omega-V)-f_S(\omega)]T_Q(\omega),
\end{eqnarray}
with a quasiparticle transmission defined as
\begin{widetext}
\begin{eqnarray}
T_Q(\omega)=4\Gamma_N\Gamma_S\theta(|\omega|-\Delta)\frac{|\omega|}{\sqrt{\omega^2-\Delta^2}}\sum_\sigma[ |G^{HF}_{11,\sigma}(\omega)|^2+ |G^{HF}_{12,\sigma}(\omega)|^2-\frac{2\Delta}{|\omega|}Re\{G^{HF}_{11,\sigma}(\omega)[G^{HF}_{12,\sigma}(\omega)]^*\}].
\end{eqnarray}
\end{widetext}
This quasiparticle contribution consists of three proccesses: 1) conventional tunneling, 2) an electron (hole) in the normal side is converted into a hole (electron) excitation in the superconducting side (branch crossing \cite{BTK}), and 3) quasiparticle transfer from the normal lead into the superconducting lead, while creating (or annihilating) a Cooper pair as an intermediate state.

\section{Analytic model for energy level repulsion}

Below we derive a simple expression which describes the level repulsion between the doublet states and the states in the continuum of the superconducting lead at small coupling $\Gamma_S\ll \Delta$, in the regions of gate voltage corresponding to a singlet ground state.

The Hamiltonian for the superconducting lead given by Eq.~(\ref{Hnu}) can be diagonalized after a Bogoliubov transformation,
\begin{equation}
H_S=\sum_{k_S\sigma}\epsilon_{k_S}\gamma^\dagger_{k_S\sigma}\gamma_{k_S\sigma},
\end{equation}
where $\gamma_{k_S\sigma}=u_{k_S} c_{k_S\sigma}+\sigma v_{k_S}c^\dagger_{-k_S-\sigma}$ are the annihilation operators of Bogoliubov quasiparticles with energy $\epsilon_{k_S}=[\Delta^2+\xi_{k_S}^2]^{1/2}$, and $u_{k_S},v_{k_S}=[(1\pm\xi_{k_S}/\epsilon_{k_S})/2]^{1/2}$ are the BCS coherence factors. The projection of the Hamiltonian $H_{QD-S}$ to a subspace of states with energy close to $\Delta$ in the leads then reads
\begin{equation}
H_{QD-S}=
\sum_{k_S\sigma}
\left(\Delta+\frac{\xi^2_{k_S}}{2\Delta}\right)\gamma^\dagger_{k_S\sigma}\gamma_{k_S
\sigma}
+
H_{QD}
+
\frac{1}{\sqrt{2}}
\sum_{k_S}
\left[
V_{k_S}
\left(\gamma^\dagger_{k_S\sigma}-\sigma\gamma_{k_S-\sigma}\right)d_{\sigma}
+
\mathrm{h.c.}
\right].
\label{eq:Hred}
\end{equation}
The sums in the r.h.s of this equation are restricted to momenta such that $|\xi_{k_S}|\ll\Delta$, and $u_{k_S},v_{k_S}\approx1/\sqrt{2}$. 

At vanishing tunnel coupling, the ground state at $\epsilon_\uparrow>0$ is the product state of the BCS ground state in the lead and the empty level in the dot, which we denote $|\emptyset \rangle$. Lowest energy excited states consist of the singly occupied level, $d^\dagger_\sigma|\emptyset\rangle$, as well as the BCS ground state filled with one Bogoliubov quasiparticle, $\gamma^\dagger_{k_S\sigma}|\emptyset\rangle$. When their energies $\epsilon_\sigma$ and $\epsilon_{k_S}$ are close and the tunnel couplings $V_{k_S}$ are finite, the discrete state on the dot and the continuum of states in the leads hybridize. Then, we may use $|\Psi_\sigma\rangle=(A d^\dagger_\sigma+\sum_{k_S} B_{k_S}\gamma^\dagger_{k_S\sigma})|\mathrm{\emptyset}\rangle$ as a variational wavefunction for the excited states with spin $\sigma$. From the set of equations
\begin{subequations}
\begin{eqnarray}
\left(E-\epsilon_\sigma\right)A&=&\frac 1{\sqrt{2}}\sum_{k_S}V_{k_S}^*B_{k_S},
\\
\left(E-\Delta-\frac{\xi_{k_S}^2}{2\Delta}\right)B_{k_S}&=&\frac 1{\sqrt{2}}V_{k_S}A,
\end{eqnarray}
\end{subequations}
that determine possible eigenenergies $E$ associated with the wavefunction $|\Psi_\sigma\rangle$, we obtain the following equation for the bound state excitation energy $\zeta_\sigma\approx E$ of state $|\sigma\rangle$:
\begin{equation}
\zeta_\sigma-\epsilon_\sigma=-\Gamma_S\sqrt{\Delta/[2(\Delta-\zeta_\sigma)]}.
\end{equation}
It yields
\begin{equation}
\zeta_\sigma\simeq
\begin{cases}
\epsilon_\sigma-\Gamma_S\sqrt{\Delta/[2(\Delta-\epsilon_\sigma)]},
&
\Delta-\epsilon_\sigma\gg (\Gamma_S^2\Delta)^{1/3},
\\
\Delta-(\Gamma_S^2\Delta/2)^{1/3},
&
\epsilon_\sigma\simeq\Delta,
\\
\Delta-\Gamma_S^2\Delta/[2(\epsilon_\sigma-\Delta)^2],
&
\epsilon_\sigma-\Delta\gg (\Gamma_S^2\Delta)^{1/3}.
\end{cases}
\label{eq:zeta}
\end{equation}
Equation (\ref{eq:zeta}) describes the anticrossing (or level repulsion) between the dot state and the BCS continuum.
Namely, a bound state forms at all values of $\epsilon_\sigma$; it gets closer to the edge of the BCS continuum -- and the splitting $|\zeta_\downarrow-\zeta_\uparrow|$ vanishes -- as $\epsilon_\sigma$ is increased. 
Note that the excitation energy of the bound state coincides with the Andreev level energy obtained from Eq.~(\ref{andreev-level}) at $\Gamma_S\ll \Delta$, in the regions near the edge of the continuum spectrum in the lead, where $\langle n_\uparrow\rangle=\langle n_\downarrow\rangle=0$ and the last term in the l.h.s gives a negligible correction in $(\Gamma_S/\Delta)^{2/3}\ll 1$.

Similarly, when $\epsilon_{-\sigma}$ is close to $-(U+\Delta)$, and the dot is doubly occupied in the ground state $d^\dagger_\uparrow d^\dagger_\downarrow|\emptyset\rangle$ at vanishing tunnel coupling, we may use 
$|\Psi_\sigma\rangle=(A d^\dagger_{\sigma}+\sum_{k_S} B_{k_S}\gamma^\dagger_{k_S\sigma}d^\dagger_\uparrow d^\dagger_\downarrow)|\emptyset\rangle$
as a variational wavefunction at finite coupling and obtain the equation
\begin{equation}
\zeta_\sigma+\epsilon_{-\sigma}+U=-\Gamma_S\sqrt{\Delta/[2(\Delta-\zeta_\sigma)]}.
\end{equation}
for $\zeta_\sigma\approx E-(2\epsilon+U)$.
It is also in correspondence with Eq.~(\ref{andreev-level}) in the regions where $\langle n_\uparrow\rangle=\langle n_\downarrow\rangle=1$.

\section{Magnetic field dependence of $\Delta$}

The magnetic field dependence of the superconducting gap, $\Delta$, was estimated from the $dI/dV$ vs $(B,V)$ measurement shown in Fig. S1a, which was taken at the center of an even valley. As discussed in the main text, the Andreev levels appear at $eV \pm\Delta$ when the system lies deep inside the singlet GS. Thus, $\Delta(B)$ is readily obtained from the evolution of the sub-gap resonances as a function of $B$ (Fig. S1b). This experimental dependence was used for obtaining the theoretical $dI/dV$ plots at finite $B$ (shown in Figs. 2 and 3 of the main text).

\begin{figure}[h]
\includegraphics[width=85mm]{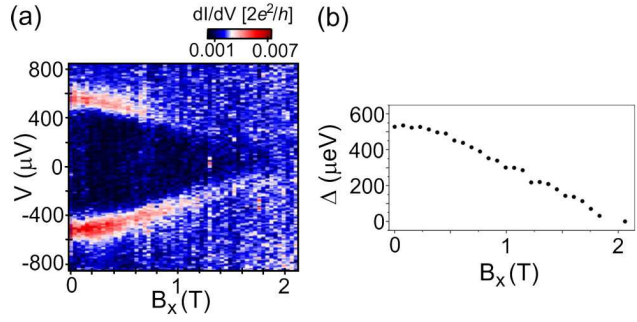}
\caption{\label{fig:epsart} (a) $dI/dV$ vs. $(B,V)$ measurement taken at the center of an even valley. (b) Superconducting gap extracted from (a) as a function of $B_x$.}
\end{figure}

\section{Characterization of second $S$-QD-$N$ device}

Here we present additional data obtained from the device discussed in Figs. 3 and 4 of the main text. The device was first characterized in the normal state (Fig. S2a). To this end, an external magnetic field $B$ = 2.5 T was applied above the critical field $B_c \approx$ 2 T, driving the vanadium electrical leads across the superconducting transition. The resulting charge stability diagram revealed a split Kondo resonance within the odd-occupied diamond (highlighted by the white arrows). Furthermore, a charging energy $U \approx$ 4.5 meV is extracted from the height of the Coulomb diamond, whereas an average tunnel coupling $<\Gamma>$ $\approx$ 0.8 meV is estimated from the FWHM of the Coulomb resonances. 

The series of $dI/dV$ vs. ($V_{pg}, V$) plots shown in Fig. S2b ($B$ = 0, 0.5 and 0.75 T) depicts the $B$-field evolution of the sub-gap Andreev levels in the second device. It is noteworthy that the data herein discussed shows a shift of $\approx$ 20 meV in the $V_{pg}$ axis when compared to the corresponding plot in the main text (Fig. 3a). This can be attributed to a local charge rearrangement in the vicinity of the QD which, due to capacitive coupling, effectively results in a small shift of the charge stability diagram. In spite of the shift, the $dI/dV$ features clearly remain unaltered. The data shown in Fig. S2 is qualitatively analogous to the situation discussed in the top (experimental) row of Fig. 2 in the main text. Specifically, at $B = 0$ (left-most panel), the sub-gap resonances cross twice the Fermi level, delimiting the boundaries between the singlet ($S$) and doublet ($D$) ground states (GSs). In agreement with the discussion in the manuscript, the Andreev levels split with increasing magnetic field when the system lies in the singlet GS.

\begin{figure}[h]
\includegraphics[width=160mm]{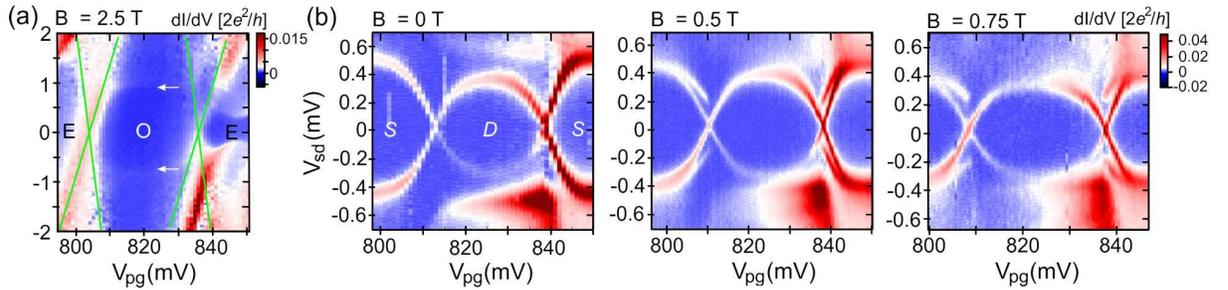}
\caption{\label{fig:epsart} (a) Characterization of second device in the normal state ($B$ = 2.5 T). The green lines are guides to the eye depicting the limits of the Coulomb diamonds with even (E) and odd (O) occupation. The white arrows highlight inelastic cotunneling steps in $dI/dV$ related to the split Kondo resonance. (b) Effect of $B$ on the Andreev levels observed in the same device. A series of $dI/dV$ vs ($V_{pg}, V$) plots taken at $B$ = 0, 0.5 and 0.75 T is shown. $S$ and $D$ indicate the regions in which the ground state is a singlet or a doublet, respectively.}
\end{figure}

\section{Supplementary data on energy level repulsion effect}

Fig. S3 contains further information concerning the energy level repulsion effect discussed in the manuscript. The right panel of Fig. S3a displays an additional $dI/dV$ vs ($B, V$) measurement taken at the position of the dashed line in the left panel, which is further away from the crossing point than position 2 in Fig. 3a of the main text. As discussed in the manuscript, the $B$-dependence of the $\zeta_{\downarrow}$ peaks deviates from the expected Zeeman splitting of the doublet state, due to the level repulsion of the $|\downarrow>$ state by the continuum of quasiparticle states above the superconducting gap. Fig. S3b demonstrates that this effect becomes more pronounced as the energy of the $\zeta_{\downarrow}$ peaks approaches $\Delta$, as expected from theory. The plotted $g$-factor values were estimated from the slopes of the $\zeta_{\uparrow,\downarrow}$ vs. $B$ data (red circles and blue triangles, respectively). The horizontal dashed line is positioned at $g \approx 5.5$, corresponding to the value estimated from the inelastic cotunneling $dI/dV$ steps in the normal state (shown in Fig. S3c). By its turn, the vertical dashed line signals the position of the singlet-doublet phase boundary. The plot clearly shows how the $g$-factor extracted from the $\zeta_{\downarrow}$ peaks is strongly suppressed when moving away from the crossing point. The values obtained from the $\zeta_{\uparrow}$ peaks, on the other hand, show no significant plunger gate dependence.

\begin{figure}[h]
\includegraphics[width=86mm]{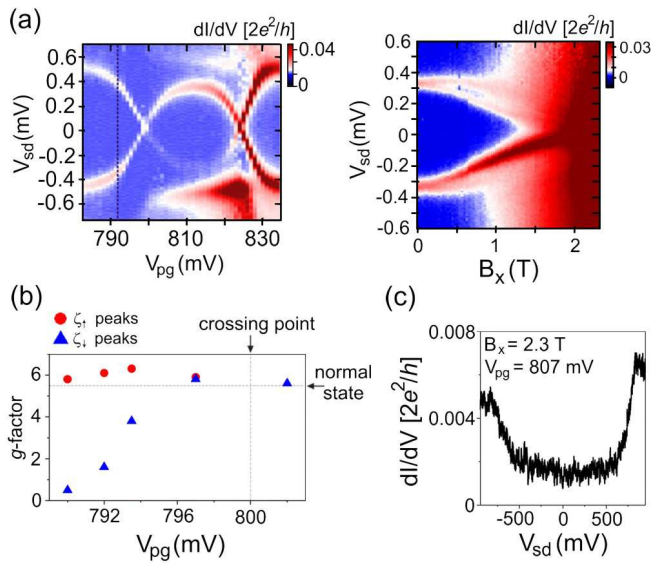}
\caption{\label{fig:epsart} (a) Right panel: $dI/dV$ vs. ($B, V$) plot taken at the position marked by the dashed line in the left panel. The measurement shown in the left panel is the same as that shown in Fig. 3a in the main text. (b) $g$-factor estimated from $\zeta_{\uparrow}$ and $\zeta_{\downarrow}$ peaks as a function of $V_{pg}$. The level repulsion effect is evidenced by the suppressed $\zeta_{\downarrow}$ $g$-factor at positions further away from the crossing point, where the $\zeta_{\downarrow}$ peak energy approaches $\Delta$. (c) Inelastic cotunneling $dI/dV$ steps resulting from the field-induced splitting of a spin-1/2 Kondo resonance. This measurement, taken at $B$ = 2.3 T, yields a normal state $g \approx 5.5$.}
\end{figure}

\section{Supplementary data on angular dependence of the Andreev levels at finite $B$}

\begin{figure}[h]
\includegraphics[width=160mm]{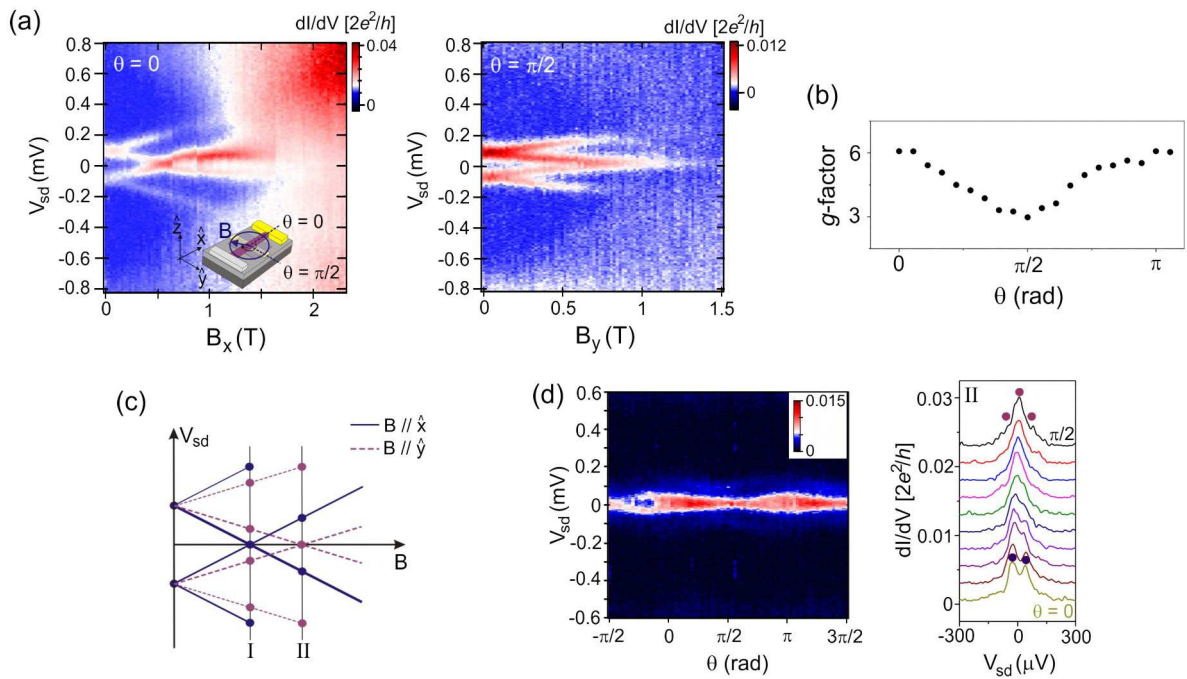}
\caption{\label{fig:epsart} (a) $B$-field dependences of the Andreev levels measured at $\theta=0$ (left panel) and $\theta=\pi/2$ (right panel). The inset in the left panel shows the orientation of the field with respect to the nanowire ($\theta =0$ corresponds to $B$ aligned parallel to NW). (b) $g$-factor anisotropy measured from the Andreev level splitting at $|B|$ = 0.6 T. (c) Schematic diagram summarizing the angular dependence behaviour. Due to the $g$-factor anisotropy, the QPT occurs at different fields for $\theta = 0$ (solid blue line) and $\pi/2$ (dashed violet line). Lines I and II highlight the $dI/dV$ features observed when $|B|$ is fixed at $|B_{QPT}^x|$ and $|B_{QPT}^y|$, respectively.
(d) Angular dependence measurement taken with $|B|=|B_{QPT}^y$ = 0.35 T. Here, the zero-bias QPT peak appears for $B$ perpendicular to the nanowire axis. The circles in the line profile plot (right panel) ascribes the measured $dI/dV$ peaks to the Andreev levels shown in (c).
}
\end{figure}

To complement the angular dependence data shown in Fig. 4b of the main text, we include here the complete $B$-field dependences measured at $\theta = 0, \pi/2$ (Fig. S4a). The plot shown in the left panel ($\theta = 0$) is analogous to the measurement shown in Fig. 4a of the main text. From this, a QPT field $B_{QPT}^{x} \approx$ 0.6 T is estimated. 
The corresponding QPT measurement at $\theta = \pi/2$ (right panel) shows the same qualitative features, however with two significant differences. The most relevant difference is that the slope of the Andreev levels is reduced, indicating a smaller $g$-factor. This results in a QPT field $B_{QPT}^{y} >$ 1 T. The $g$-factor anisotropy is highlighted in Fig. S4b, where the data points were extracted from the angular dependence of $\zeta_{\uparrow}$ and $\zeta_{\downarrow}$ at a field magnitude $|B|$ = 0.6 T (data shown in Fig. 4b of the main text). The second difference between the two data plots in Fig. S4a is in the value of the respective critical fields: $B_c^x\approx 2$T against $B_c^x\approx 1.5$T.  

The scheme depicted in Fig. S4c summarizes the evolution of the Andreev levels for the longitudinal (solid line) and perpendicular (dashed line) field directions, as deduced from the data in panel (a). For a field amplitude $|B|= |B_{QPT}^x| =$ 0.6 T, a zero-bias peak (ZBP) is observed at $\theta$ = 0 (or $\theta = \pi$) in correspondence of the blue dot along line I. At $\theta=\pi/2$ (or $\theta=3\pi/2$), the ZBP is split into two peaks in correspondence of the violet dots along line I. This is indeed the behaviour observed in Fig. 4b of the main text. The scenario is reversed when the field magnitude $|B|$ is fixed at $|B_{QPT}^y|$ (as indicated by line II). Unfortunately, the QPT in the perpendicular direction occurs near the closing of the gap (Fig. S4a). To better visualize the angular dependence of the Andreev levels in the case $|B|= B_{QPT}^y$, the $V_{pg}$ position was moved closer to the singlet-doublet phase boundary, so that the Andreev levels are closer to the Fermi level at $B=0$. Fig. S4d shows the angular dependence measurement taken at $|B|$ = 0.35 T. In this case a ZBP is observed for $\theta=\pi/2$ (or $\theta=3\pi/2$), in correspondence of the violet dot along line II, while a split peak is observed for $\theta=0$ (or $\theta=\pi$), as denoted by the pair of blue dots along line II.

\section{Relevance to tunnel spectroscopy experiments aiming at the observation of Majorana fermions}

Following recent theoretical proposals \cite{Lutchyn, Oreg}, the past years have seen intense experimental efforts for the realization and detection of Majorana fermions (MFs) in hybrid superconductor-semiconductor nanowire devices. These exotic quasiparticles are predicted to arise in hybrid devices consisting of a semiconductor nanowire with strong spin-orbit coupling (\textit{e.g.}, InSb, InAs) coupled to an $s$-wave superconductor, and in the presence of an external magnetic field, $B$, applied perpendicular to the Rashba spin-orbit field, $B_{SO}$. When the Zeeman energy, $E_{Z}=g\mu_{B}B$, exceeds the critical value  $2\sqrt{\Delta_{ind}^{2}+\mu^2}$, where 
$\Delta_{ind}$ is the superconducting gap induced in the NW by the proximity effect and $\mu$ is the chemical potential, the NW section underneath the superconductor undergoes a transition into a topological superconducting phase, and zero-energy MF states are formed at its edges. 
So far, the experimental hunt for MFs in hybrid devices has predominantly relied on dc transport experiments aiming at the detection of these zero-energy states through ZBPs in the differential conductance \cite{Mourik, Heiblum, Deng, Finck, Churchill}. 
In this relatively simple approach, it is essential to understand and rule out all other physical effects that could give rise to ZBPs (\textit{e.g.}, reflectionless tunneling \cite{vanWees}, weak antilocalization \cite{WAL}, Kondo effect \cite{Goldhaber}, etc.). 
Here we extend our discussion on the ZBPs resulting from Andreev levels crossing the Fermi energy under the action of an applied magnetic field. We show that this physical mechanism is particularly relevant since it can produce experimental signatures that can resemble very much those expected from MFs. We aim to highlight such signatures, make comparisons to existing experiments, and provide guidelines to distinguish ZBPs stemming from zero-energy Andreev levels from those due to MFs. Our work deals with QDs with on-site Coulomb repulsion, rather than non-interacting one-dimensional structures, which are required by the proposals on MFs. Yet we wish to emphasize that our conclusions may still hold relevance to existing experiments where, to our view, interactions cannot be ruled out with certainty (we note that the reported ZBPs have amplitudes below the conductance quantum, and that electron localization may arise even at the level of individual barriers formed by a local charge depletion of the nanowire).

ZBPs due to MFs are predicted to follow a characteristic magnetic-field dependence. In superconductor-nanowire devices, a MF ZBP is expected to emerge only for finite fields perpendicular to $B_{SO}$ (due to symmetry arguments, this latter should lie most likely in the device plane perpendicular to the nanowire axis). Therefore, logically, the MF ZBP should disappear when a rotation brings the magnetic field parallel to $B_{SO}$. For field directions perpendicular to $B_{SO}$, the ZBP should be stable against further increases in the field magnitude, provided the nanowire is sufficiently long. In this case, the two MF modes located at the ends of the topological nanowire do not overlap, and consequently, remain at zero energy. By contrast, shorter nanowires are expected to display a sizable splitting of the ZBP due to the hybridization of the MF modes. It is predicted that this splitting should be modulated in an oscillatory fashion by variations in the magnitude of the magnetic field \cite{Prada,Loss,Sarma}.  

Let us now discuss to what extent the just mentioned fingerprints of MF ZBPs can be found in Andreev-level ZBPs.

\begin{figure}[h]
\includegraphics[width=160mm]{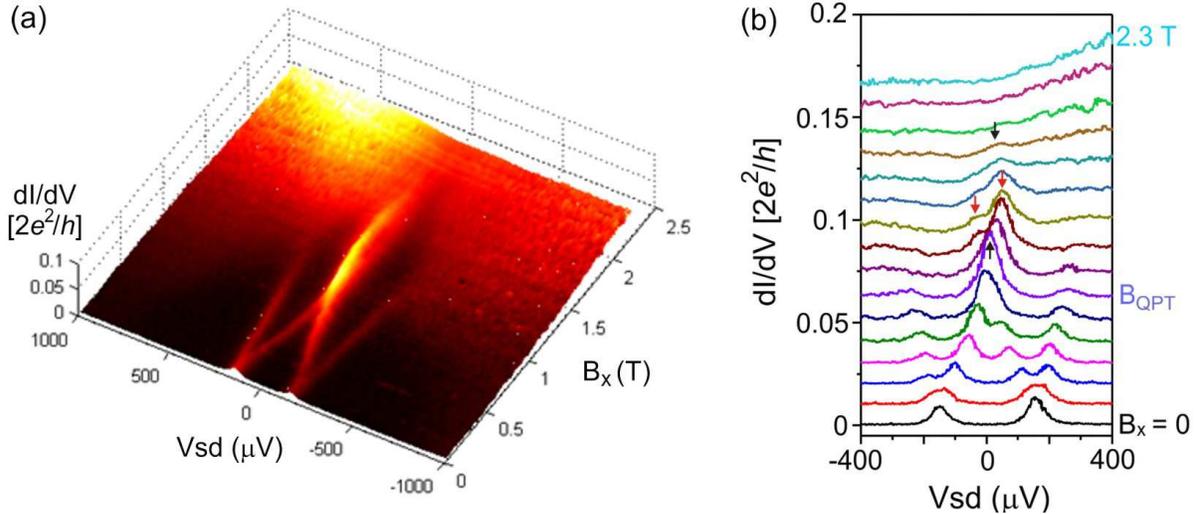}
\caption{\label{fig:epsart} (a) $dI/dV(B_x,V)$ taken at a position slightly further from the singlet-doublet crossing as that shown in Fig. 4a (main text). As a result, the QPT ZBP appears at a larger magnetic field ($B^{QPT}_x \approx$ 0.9 T). (b) $dI/dV(V)$ line profiles of the dataset shown in (a). With increasing field, a black arrow first highlights the QPT ZBP ($B = B_{QPT}$). A second black arrow indicates another ZBP which results from the "squeezing" of $\zeta$ peaks (marked by red arrows) with the closing of the gap. This re-emergence of the ZBP with increasing field qualitatively mimicks the oscillatory hybridization of MFs.
}
\end{figure}


\subsection{Emergence of Andreev-level ZBPs at finite field and their stability against $B$ variations.}

As shown in Fig. 4 (main text), the observed ZBP persists over a field range $\Delta B_x \approx$ 150 mT. This range is consistent with that expected from the finite width $w$ of the Andreev level crossing the Fermi energy, i.e. $\Delta B \approx w/(|g|\mu_B)$. Interestingly, the ZBP reported by Das {\it et al.} \cite{Heiblum} (hybrid devices based on InAs NWs coupled to Al superconducting electrodes), displays features qualitatively very similar to those of Fig. 4 (main text). To emphasize this similarity we show in Fig. S5a a data set of the same type plotted on a three-dimensional color scale similar to the one of Fig. 4a in ref. \citenum{Heiblum}. Thus, let us discuss how far an Andreev-level picture can be fitted to the experimental results of ref. \citenum{Heiblum}.
 
In Fig. 4a of ref. \citenum {Heiblum}, two $dI/dV$ peaks at $V_{sd} = \pm 45$ $\mu$eV are observed at zero magnetic field. These peaks are interpreted as the edges of the induced superconducting gap, $\Delta_{ind}$. The fact that the two peaks approach each other as a result of an applied magnetic field is interpreted as the closure of this gap. Here we consider a different scenario where the two peaks (at $B=0$) correspond to Andreev levels associated with the transition from a singlet ground-state to a doublet excited state. As discussed in the main text, increasing the field results in the Zeeman splitting of these levels. By carefully looking at Fig. 4a of ref. \citenum{Heiblum}, a splitting may indeed be seen in the data although the split peak moving to higher energies disappears quickly into the edges of the superconducting gap. This seems even clearer in Fig. S12 of the same reference. On the other hand, no splitting is expected in the alternative scenario of a field-driven closure of the induced superconducting gap as confirmed by the numerical calculations in Fig. 4d of ref. \citenum{Heiblum}.
Thus, in the Andreev-level picture, the induced gap closure can be interpreted instead as the lowest-energy Zeeman-split level, $\zeta_\uparrow$, evolving linearly towards the Fermi energy. 

In Fig. 4a of ref. \citenum{Heiblum}, the ZBP appears to extend approximately from 40 to 70 mT. In the Andreev level picture, this corresponds to having $\zeta_{\uparrow}$ = 0 at B= 55 mT. Given that $\zeta = 45 \mu$eV for B=0, we estimate $|g| \approx 14$, a value pretty close to the electron g-factor in bulk InAs. From the energy width of the ZBP we estimate $w \approx 20 \mu$eV. Based on these values, the expected field extension of the ZBP is $\Delta B \approx w/(|g|\mu_B) \approx 25$ mT, which is consistent with the field range estimated from the experimental data. Above 70 mT, the ZBP splits again. As opposed to an interpretation based on coupled MF edge states, this splitting can be readily understood as a result of the fact that increasing $B$ stabilizes the $|\uparrow\rangle$ ground state leading to a sizable and growing excitation energy to the $|-\rangle$ state.

Based on the analysis above, a Zeeman-split Andreev level crossing the Fermi energy results in a ZBP with relatively limited extension, directly proportional to the ratio between its finite life-time broadening ($w$) and the g-factor. This cannot account for the large magnetic-field robustness of the ZBPs measured in hybrid devices based on InSb nanowires coupled to Nb-based superconducting electrodes. For instance, Mourik et al. \cite{Mourik} reported ZBPs extending over field ranges of several hundred mT, which is quite striking if one considers that g-factors in InSb NW are at least a few times larger than in InAs NWs. As we pointed out in the main text, however, the level repulsion between Andreev levels and the continuum of quasiparticle states can result in a significant stretching of the ZBP. In principle, a similar effect may be expected from other Andreev levels within the superconducting gap, as suggested by recent theoretical works \cite{Prada, Loss, LiuPotter, Brouwer}. The data in the left panel of Fig. 3c (main text), provide an experimental demonstration of this effect as confirmed by the numerical simulations in the right panel of the same figure. Following the $|-\rangle \rightarrow |\uparrow\rangle$ quantum-phase transition, i.e.  for $B > B_{QPT}$, the $\zeta$ peaks corresponding to the excitation from $|\uparrow\rangle$ to $|-\rangle$ are repelled by the quasiparticle states at the gap edge. As a result, the split peaks remain squeezed to zero-bias, thus stretching the ZBP up to a several hundred mT, i.e. well beyond the range expected from the $w/g$ ratio. Having said so, the Andreev-level ZBPs observed in the present work exhibit a splitting at lower and, in cases like Fig. S5, higher fields. This characteristic is apparently absent in the ZBPs reported for InSb nanowires contacted by Nb-based superconducting electrodes \cite{Mourik,Churchill}. This dissimilarity seems to rule out an interpretation of those ZBPs within the simple picture of an Andreev-level pair at the Fermi energy.

\subsection{Dependence of the Andreev-level ZBPs on the field angle.}

The suppression of the ZBP when the external field is applied parallel to the spin-orbit field has been considered as one of the strongest indications in favor of a MF interpretation. We note that most of the reported experiments present their angular dependence data by fixing $B$ at a position where the ZBP is visible, and then performing a field rotation at constant field magnitude. As mentioned in the main text, by carrying out our measurement of Andreev-level ZBPs in a similar way, we were able to recover the same qualitative features expected for MFs. As the angle dependence of Andreev-level ZBPs originates from $g$-factor anisotropy, however, a ZBP is also expected for $B \parallel B_{SO}$, yet at a different field magnitude (see Fig. S4). Thus, we argue that carrying out a full $B$-dependence also for $B \parallel B_{SO}$ (such as in Fig. S4a), and not simply a field rotation at constant field magnitude, is a useful control experiment to discriminate ZBPs due to MFs from those due to Andreev-levels. 



\subsection{Splitting of the ZBP and its magnetic-field induced oscillations.}

Recently, a few experimental studies have reported either the splitting of the ZBP with increasing $B$ \cite{Heiblum}, or a ZBP that appears and vanishes a few times over a large B sweep \cite{Finck,Churchill}. These experimental observations have been interpreted as possible evidences of the coupling between MFs at the opposite edges of a nanowire segment with induced topological superconductivity. In the main text we have shown that the Andreev-level ZBP appears at $B = B_{QPT}$, where $\zeta_\uparrow = 0$ (Fig. 4a). Either above or below this field, a splitting of the ZBP is observed (Fig. 4a, Fig.S5). Therefore, in the simplest case of a single-level QD considered here, only one ZBP is expected, which is centered around $B = B_{QPT}$. Still, due to the field-induced suppression of the superconducing gap, the split $\zeta$ peaks for $B > B_{QPT}$ converge back to zero-energy, an effect that can resemble like a re-emergence of a ZBP when $B$ approaches to the critical field. (Fig. S5b). In case of a system with more than two Andreev levels in the superconducting gap, i.e., different than the small QD limit considered here, a ZBP may appear and disappear multiple times over a large field range due to different Andreev levels crossing the Fermi energy at different magnetic fields. This scenario has been discussed in recent theoretical works \cite{Prada,LiuPotter}.